\documentclass{PoS}                              % Documentstyle for PoS submission

\usepackage[latin1]{inputenc}                    % Codepage latin1
\usepackage{graphicx}                            % Graphics package
\usepackage{latexsym}                            % Load additional symbols
\usepackage{amsfonts}                            % Use AMS fonts,
\usepackage{amssymb}                             % symbols,
\usepackage{amsmath}                             % and definitions
\usepackage{dcolumn}                             % Align table columns
\usepackage{bm}                                  % Bold math

\graphicspath{{.}{Graphics/}{Figures/}}          % Path for graphics
\newcommand{\imag}{\mbox{i}}                     % Roman imaginary unit
\newcommand{\msbar}{$\overline{\mbox{\rm MS}}$}  % MS-bar
\title{Distribution Amplitudes of Pseudoscalar Mesons}

\ShortTitle{Meson distribution amplitudes}

\author{V.~M.~Braun$^a$, M.~G{\"o}ckeler$^a$, R.~Horsley$^b$,
  H.~Perlt$^c$, D.~Pleiter$^d$, P.~E.~L.~Rakow$^e$,
  G.~Schierholz$^{df}$, A.~Schiller$^c$,
  \speaker{W.~Schroers}$\,^d$,
  H.~St{\"u}ben$^g$, and J.~M.~Zanotti$^b$\\
\centerline{[QCDSF/UKQCD collaboration]} \\
\llap{$^a$}Institut f\"ur Theoretische Physik, Universit\"at
Regensburg, 93040 Regensburg, Germany\\
\llap{$^b$}School of Physics, University of Edinburgh,
  Edinburgh EH9 3JZ, UK\\
\llap{$^c$}Institut f{\"u}r Theoretische Physik,
  Universit{\"a}t Leipzig, 04109 Leipzig, Germany\\
\llap{$^d$}John von Neumann-Institut f\"ur Computing NIC /
  DESY, 15738  Zeuthen, Germany\\
\llap{$^e$}Theoretical Physics Division, Department of Mathematical
  Sciences, University of Liverpool, Liverpool L69 3BX, UK\\
\llap{$^f$}Deutsches Elektronen-Synchrotron DESY,
  22603 Hamburg, Germany\\
\llap{$^g$}Konrad-Zuse-Zentrum f\"ur Informationstechnik Berlin, 14195
  Berlin, Germany}

\abstract{We present results for the first two moments of the
  distribution amplitudes of pseudoscalar mesons. Using two flavors of
  non-perturbatively improved clover fermions and non-perturbative
  renormalization of the matrix elements we perform both chiral and
  continuum extrapolations and compare with recent results from models
  and experiments. \\ \\ DESY 06-170 \\ Edinburgh 2006/27
\\}

\PACS{12.88.Gc, 14.40.Aq}

\FullConference{XXIV International Symposium on Lattice Field Theory\\
                 July 23-28 2006\\
                 Tucson Arizona, US}

\begin{document}

% --------------------------------------------------------------------------
%
%  Introduction
%
% --------------------------------------------------------------------------

\section{Introduction\label{sec:introduction}}
Hadronic wave functions are of crucial importance when describing
exclusive and semi-ex\-clu\-si\-ve reactions~\cite{Brodsky:1989pv}.
For detailed references and applications consult~\cite{Braun:2006dg}.
Distribution amplitudes (DAs) are related to the hadron's
Bethe-Salpeter wave function, $\phi(x,k_\perp)$, by an integral over
transverse momenta.  For the leading twist meson-DAs we have
\begin{equation}
  \label{eq:dadef}
  \phi(x,\mu^2) = Z_2(\mu^2)\int_{\vert k_\perp\vert<\mu} d^2k\,
  \phi(x,k_\perp)\,,
\end{equation}
where $x$ is the quark longitudinal momentum fraction, $Z_2$ the
renormalization factor (in the light-cone gauge) for the quark-field
operators in the wave-function, and $\mu$ denotes the renormalization
scale. In this presentation we quote all numbers with a scale
$\mu^2=4\,$ GeV$^2$ in the \msbar-scheme.

It is convenient to rescale $\xi=2x-1$. In the following we use
$\phi(\xi)$ to describe any pseudoscalar meson, $\phi_\pi(\xi)$ to
refer to the pion, and $\phi_K(\xi)$ to denote the kaon. Furthermore,
it is common to expand DAs into their Gegenbauer moments and quote the
expansion coefficients, $a_i$, at a given renormalization scale as a
parameterization of DAs,
\begin{equation}
  \label{eq:daexpansion}
  \phi(\xi,\mu^2) = \frac{3}{4}(1-\xi^2)\left( 1+\sum_{n=1}^\infty
  a_n(\mu^2) C_n^{3/2}(\xi) \right)\,.
\end{equation}
The zeroth moment is normalized to unity, $\int_{-1}^1d\xi
\phi(\xi,\mu^2)=1$, at any energy scale $\mu^2$. From renormalization
group arguments we find that
\[ \phi(\xi,\mu^2\to\infty) = \phi_{\mbox{\tiny as}}(\xi) =
\frac{3}{4}(1-\xi^2)\,. \] Taking the $u$- and $d$-quarks to be
degenerate, $G$-parity implies that the pion DA is an even function of
$\xi$ and hence all odd moments vanish, i.e., $a^\pi_{2n+1}=0$.

Recently, we have computed the first moments of meson distribution
amplitudes~\cite{Braun:2006dg} on the lattice. Independently, a
calculation of the first moment of the kaon distribution amplitude has
appeared which uses a different discretization scheme and different
working points~\cite{Boyle:2006pw}. In this contribution we present
our calculation of the first non-vanishing moment of the pion DA,
$a^\pi_2$, and the first two moments of the kaon DA, $a^K_1$ and
$a^K_2$. We compare our results to previous estimates from sum rules
and experiment and discuss the implications of our lattice
computation.

% --------------------------------------------------------------------------
%
%  Lattice calculation
%
% --------------------------------------------------------------------------

\section{Lattice calculation\label{sec:lattice-calculation}}
On the lattice, one has to perform the light-cone operator product
expansion to find a relation between local operators and moments of
DAs w.r.t.~$\xi$. To be specific, one has
\begin{eqnarray}
  \label{eq:locopmom}
  \nonumber
  \langle\xi^n\rangle(\mu^2) &=& \displaystyle\int_{-1}^1
  d\xi\,\xi^n\, \phi(\xi,\mu^2)\,, \\
  \langle\Omega\vert{\cal O}_{\lbrace\nu_0\dots\nu_n\rbrace}(0) \vert
  \mbox{PS}\rangle &=& \imag f_{\mbox{\tiny PS}} p_{\lbrace\nu_0}\dots
  p_{\nu_n\rbrace} \langle\xi^n\rangle\,, \\ \nonumber
  {\cal O}_{\nu_0\dots\nu_n}(0) &=& \imag^n\, \bar{q}(0)
  \gamma_{\,\,\,\,\nu_0}\gamma_5 \stackrel{\leftrightarrow}{D}_{\nu_1}\dots
  \stackrel{\leftrightarrow}{D}_{\nu_n} u(0)\,,
\end{eqnarray}
where ``PS'' refers to either the pion or the kaon, and $q$ can be
either a $d$- or an $s$-quark. $\stackrel{\leftrightarrow}{D}$ is the
covariant derivative and $\lbrace\dots\rbrace$ denotes the
symmetrization of indices and the subtraction of traces.

The matrix element, Eq.~\eqref{eq:locopmom}, can be obtained from an
appropriate ratio of two-point functions, consult~\cite{Braun:2006dg}
for details. For the first moment of pseudoscalar mesons containing
non-de\-ge\-ne\-ra\-te mass quarks we use the following two operators
\begin{equation}
  \label{eq:o41a}
  {\cal O}^a_{41} = {\cal O}_{\lbrace 41\rbrace}\,,\quad
  {\cal O}^b_{44} = {\cal O}_{\lbrace 44\rbrace} - \frac{1}{3}\left(
  {\cal O}_{\lbrace 11\rbrace} + {\cal O}_{\lbrace 22\rbrace} + {\cal
  O}_{\lbrace 33\rbrace}\right)\,.
\end{equation}
For ${\cal O}^a_{41}$ we have used the external momentum
$\vec{p}=(2\pi/L,0,0)$ --- with $L$ being the spatial lattice size ---
and the corresponding rotated momenta, for ${\cal O}^b_{44}$ we have
taken $\vec{p}=\vec{0}$. For the second moment we have only employed
${\cal O}_{412} = {\cal O}_{\lbrace 412\rbrace}$ with external
momentum $\vec{p}=(2\pi/L,2\pi/L,0)$.

We have generated our gauge field configurations using the Wilson
gauge action and two flavors of dynamical, non-perturbatively improved
clover fermions. For four different values of $\beta=5.20, 5.25,
5.29$, and $5.40$ and up to four different $\kappa$ values per $\beta$
we have produced ${\cal O}(2000-8000)$ trajectories. Lattice spacings
and spatial volumes vary between $0.075-0.123$ fm and
$(1.5-2.2\,\mbox{fm})^3$, respectively. The scale has been set using a
Sommer parameter of $r_0=0.467$ fm. For further details
see~\cite{Gockeler:2005rv}. Correlation functions are calculated every
$10$ HMC trajectories using four different locations of the fermion
source. By applying binning we obtain an effective distance of $20$
trajectories. We observe that the bin size has little effect on the
error, indicating that residual autocorrelations are small.

The matching between the lattice results and the \msbar-scheme has
been done non-perturbatively. The renormalization procedure has been
detailed in Refs.~\cite{Gockeler:2006nb,Gockeler:2004wp} and will be
discussed further in a forthcoming publication.

% --------------------------------------------------------------------------
%
%  Numerical results
%
% --------------------------------------------------------------------------

\section{Numerical results\label{sec:numerical-results}}
\subsection{Mass-degenerate quarks\label{sec:mass-degen-quarks}}
For mass-degenerate quarks, i.e., the pion, we compute
$\langle\xi^2\rangle_\pi(\mu^2=4\,\mbox{GeV}^2)$, and from this
$a_2^\pi(4\,\mbox{GeV}^2)$. To obtain the result at the physical pion
mass we first perform a linear chiral extrapolation of
$\langle\xi^2\rangle$ to the physical pion mass at each fixed value of
$\beta$.
\begin{figure}[htb]
  \begin{center}
    \includegraphics[scale=0.4,angle=270]{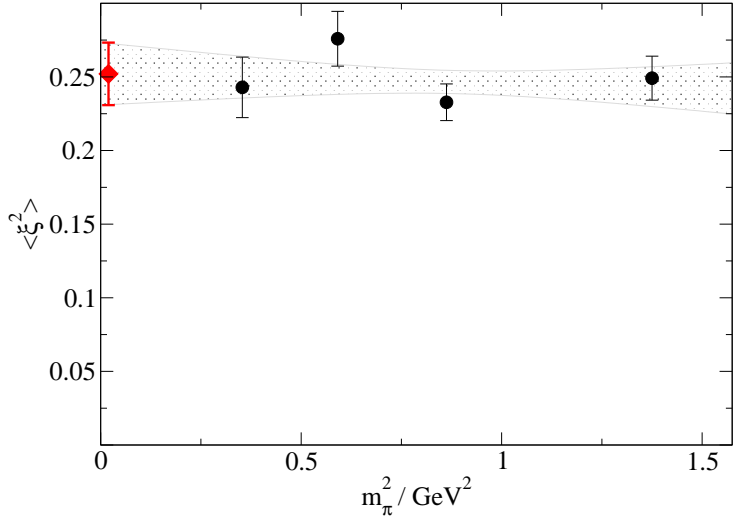}
  \end{center}
  \caption{Pion mass dependence of the second moment of the pion
    distribution amplitude, $\langle\xi^2\rangle_\pi$, at
    $\beta=5.29$, for all four values of $\kappa_{\mbox{\tiny
        sea}}=\kappa_{\mbox{\tiny val}}$ from ${\cal O}_{412}$.}
  \label{fig:xi2b529}
\end{figure}
Figure~\ref{fig:xi2b529} shows $\langle\xi^2\rangle$ obtained from
${\cal O}_{412}$ as a function of the pion mass at a fixed value of
$\beta=5.29$ with $\kappa_{\mbox{\tiny sea}}=\kappa_{\mbox{\tiny
    val}}$. The dependence on the pion mass turns out to be very weak.
Using chiral perturbation theory in the continuum it has been shown
\cite{Chen:2003fp} that for small pion masses the leading logarithmic
contribution can be absorbed in the pseudoscalar decay constant,
$f_\pi$. Hence, for small values of the pion mass we expect the
dependence on $m_\pi$ to be rather flat.  However, a precise matching
of lattice results and chiral perturbation theory similar to what has
been done in \cite{Edwards:2005ym} for the nucleon axial coupling,
$g_A$, is still to be performed.

\begin{figure}[htb]
  \begin{center}
    \includegraphics[scale=0.4,angle=270]{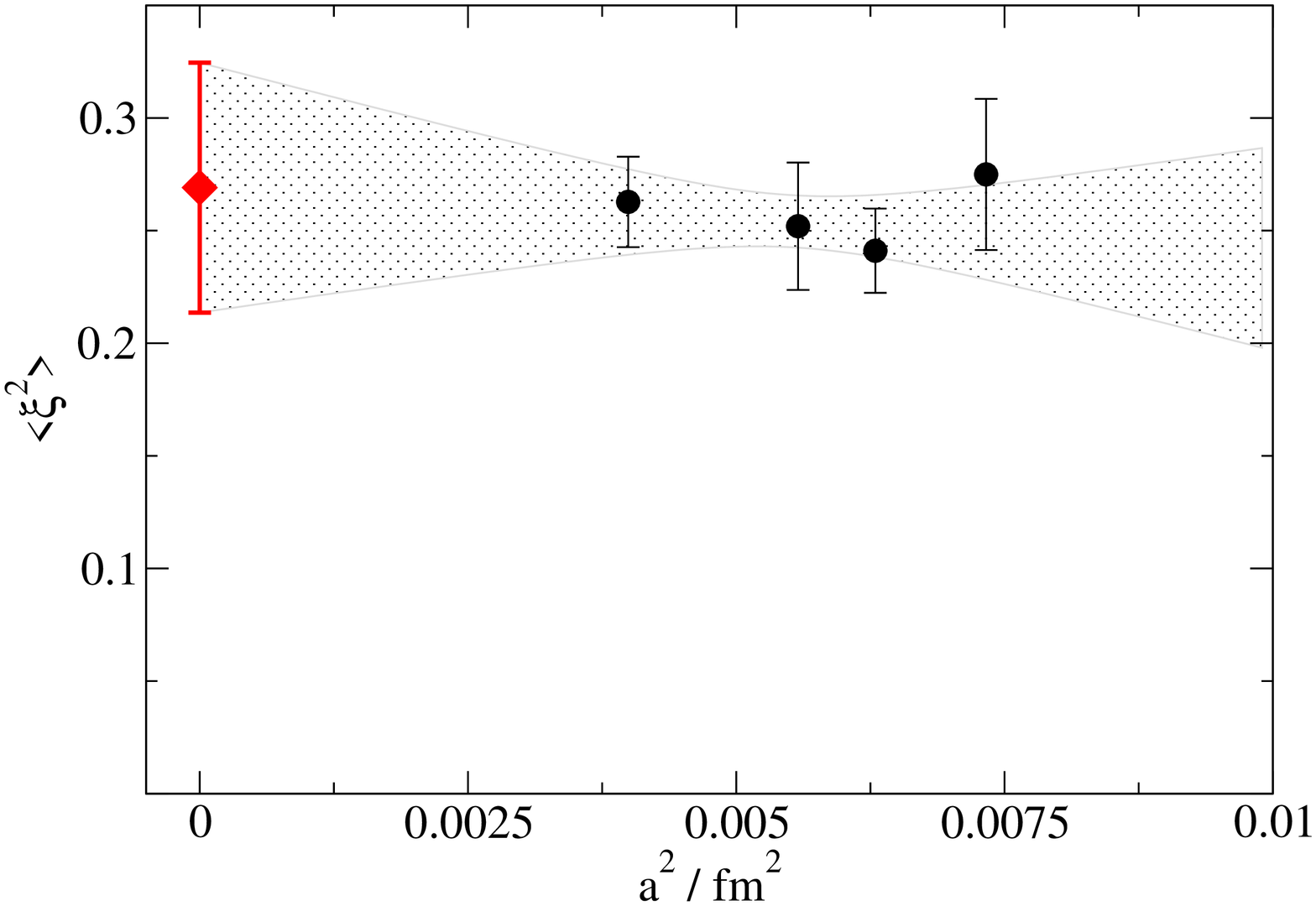}
  \end{center}
  \caption{Continuum extrapolation of $\langle\xi^2\rangle_\pi$ for
    all values of $\beta$ for ${\cal O}_{412}$.}
  \label{fig:xi2vsa}
\end{figure}
Figure~\ref{fig:xi2vsa} shows the continuum extrapolation of the
results from all values of $\beta$ at the physical pion mass for
${\cal O}_{412}$. Scaling violations seem to be small, introducing an
uncertainty of roughly $6\%$. The continuum result reads
\begin{equation}
  \label{eq:xi2pi}
  \langle\xi^2\rangle_\pi(\mu^2=4\,\mbox{GeV}^2) = 0.269(39)\,,\quad
  a_2^\pi(4\,\mbox{GeV}^2) = 0.201(114)\,.
\end{equation}
It is in agreement with the older quenched result
$\langle\xi^2\rangle_\pi(4\,\mbox{GeV}^2) =
0.286(49)^{+0.030}_{-0.013}$ from \cite{DelDebbio:2002mq}.  It is
larger than the asymptotic value,
$\langle\xi^2\rangle_\pi(\mu^2\to\infty)=0.2$, indicating that the
commonly adopted asymptotic ansatz at this energy scale may not be
justified quantitatively.

The coefficient $a_2^\pi$ has been under experimental scrutiny
recently. The decay of neutral pions into two photons contains
information on the pion wavefunction. However, as has been reported in
\cite{Diehl:2001dg} the leading twist expression is insufficient to
determine $a_2^\pi$ unambiguously. QCD sum rules,
e.g.~\cite{Chernyak:1983ej,Ball:2006wn,Bakulev:2004ar} have found that
$a_2$ is indeed positive and the more recent calculations in this
approach~\cite{Ball:2006wn,Bakulev:2004ar} are compatible with our
result also in magnitude. However, the contribution of $a^\pi_4$ is
expected to be non-negligible in modeling the pion DA\@. This makes an
independent lattice analysis of this quantity particularly
interesting.

\subsection{Mass non-degenerate quarks\label{sec:mass-non-degenerate}}
In case of two quarks with distinct masses, the odd moments will no
longer vanish. Since we are primarily interested in the light
pseudoscalar mesons we tune the parameters and extrapolations in such
a manner that the light quark (corresponding to $\kappa_{\mbox{\tiny
    sea}}$) has the correct mass to reproduce a pion if the quarks
were degenerate and the heavy quark (corresponding to
$\kappa_{\mbox{\tiny val}}$) then reproduces the kaon mass. Since we
require a large set of different valence masses, we have restricted
ourselves to the four data sets at $\beta=5.29$ only. These have
$\kappa_{\mbox{\tiny sea}}=0.1340, 0.1350, 0.1355,$ and $0.1359$. The
systematic error due to the continuum extrapolation is estimated from
the extrapolation in Fig.~\ref{fig:xi2vsa} to be about $6\%$.

\begin{figure}[htb]
  \begin{center}
    \includegraphics[scale=0.4,angle=270]{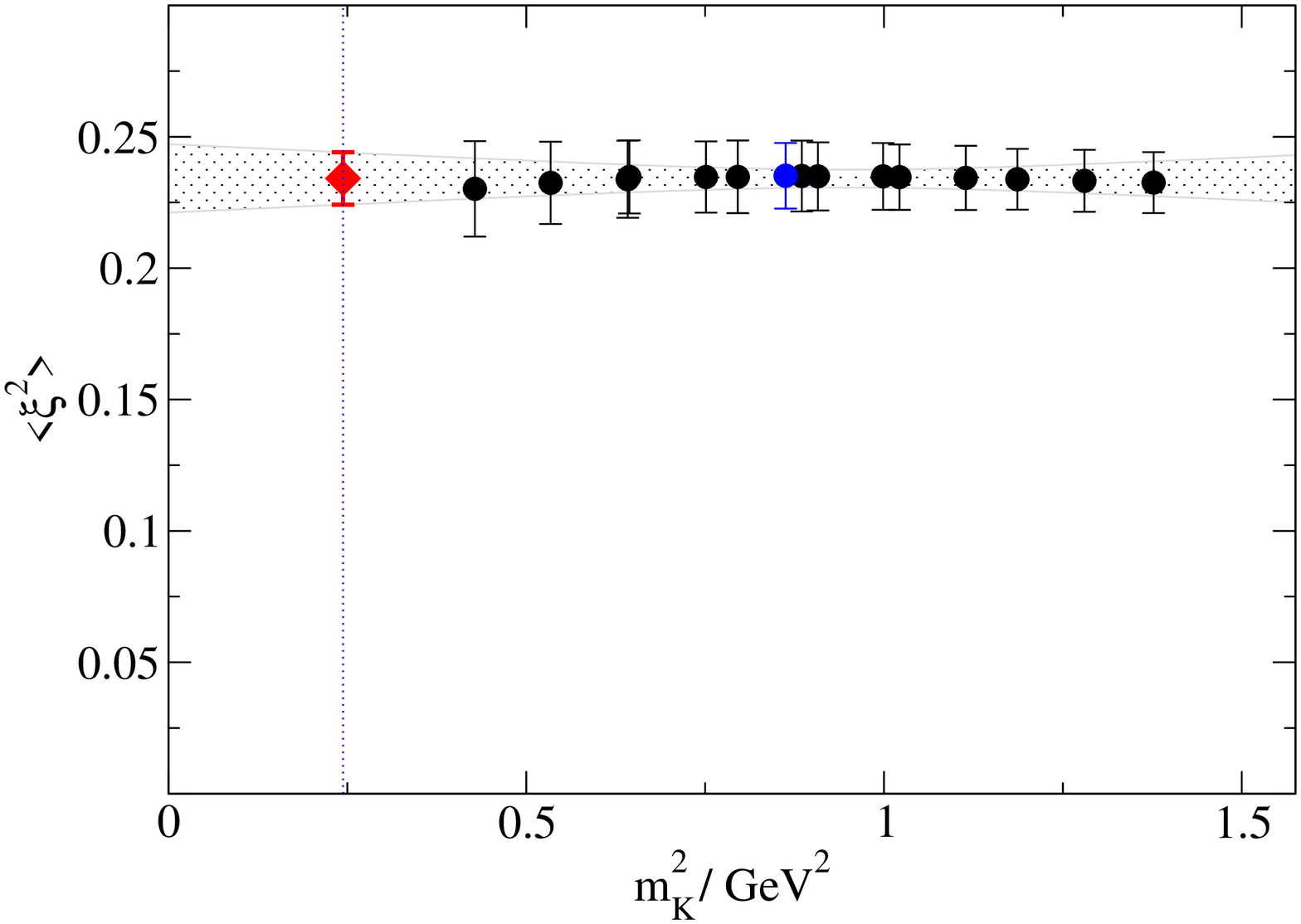}
  \end{center}
  \caption{Extrapolation of the second moment of the kaon distribution
    amplitude, $\langle\xi^2\rangle_K$ at $\mu^2=4\,\mbox{GeV}^2$, at
    $\beta=5.29$, $\kappa_{\mbox{\tiny sea}}=\kappa_{\mbox{\tiny
        light}}=0.1359$ and several values of $\kappa_{\mbox{\tiny
        val}}=\kappa_{\mbox{\tiny heavy}}$. The extrapolation to the
    physical value is shown at the vertical line. The blue data point
    indicates the degenerate mass case, $\kappa_{\mbox{\tiny
        heavy}}=\kappa_{\mbox{\tiny light}}$.}
  \label{fig:xi2vsmK}
\end{figure}
A sample plot for the second moment of the kaon distribution amplitude
can be seen in Fig.~\ref{fig:xi2vsmK} for the working point
$\beta=5.29$, $\kappa_{\mbox{\tiny sea}}=0.1359$. Several different
values have been chosen for the heavy quark mass. The vertical line
indicates the physical kaon mass and the data point at this location
has been taken from a linear extrapolation.

The data from all four samples of gauge fields at $\beta=5.29$ has
been combined and linearly extrapolated to the physical values of
$m_K$ and $m_\pi$ in a global fit, see \cite{Braun:2006dg} for
details. The resulting value is
\begin{equation}
  \label{eq:xi2Kres}
  \langle\xi^2\rangle_K(4\,\mbox{GeV}^2) = 0.260(6)(16)\,,\quad
  a_2^K(4\,\mbox{GeV}^2) = 0.175(18)(47)\,.
\end{equation}
The first error is of statistical origin while the second one
corresponds to a scaling error of $6\%$. The result \eqref{eq:xi2Kres}
corresponds to a ratio of
$\langle\xi^2\rangle_K/\langle\xi^2\rangle_\pi\simeq 1$. In the
literature, there has been the debate between \cite{Chernyak:1983ej}
predicting a ratio of $a_2^K/a_2^\pi\sim 0.59\pm 0.04$ and
\cite{Ball:2006wn,Khodjamirian:2004ga} predicting a ratio of about
$\sim 1$. Our finding clearly favors the latter results.

Of particular interest is the first moment of the kaon DA,
$\langle\xi\rangle_K$, and, hence, also $a_1^K$.
\begin{figure}[htb]
  \begin{center}
    \includegraphics[scale=0.4,angle=270]{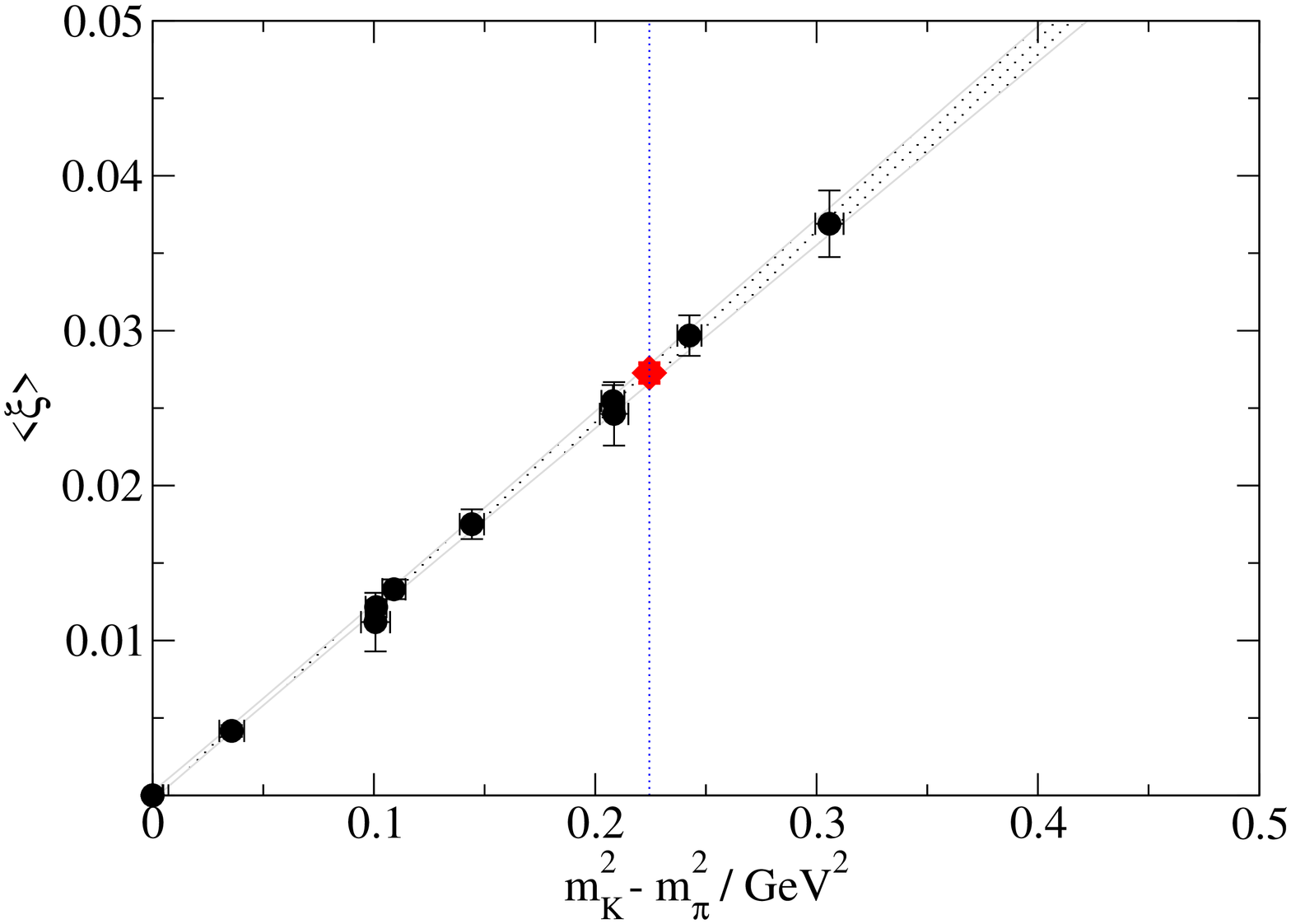}
  \end{center}
  \caption{Interpolation of the first moment of the kaon DA,
    $\langle\xi\rangle_K$ at $\beta=5.29$, $\kappa_{\mbox{\tiny
    sea}}=0.1359$ as obtained from ${\cal O}^b_{44}$. The physical
    point corresponding is indicated by the vertical line, the value
    obtained by a linear interpolation.}
  \label{fig:xivsmpsd}
\end{figure}
Figure~\ref{fig:xivsmpsd} shows the interpolation of
$\langle\xi\rangle$ obtained from ${\cal O}^b_{44}$ as a function of
the mass difference $m_K^2-m_\pi^2$ for one sample of gauge fields.
Note that this is one of the few cases where an interpolation
suffices. We now have a result at a specific pion mass determined by
$\kappa_{\mbox{\tiny sea}} = \kappa_{\mbox{\tiny light}}$. After
repeating this procedure on our other data-sets with $\beta=5.29$, we
extrapolated the results linearly in the sea quark mass to the
physical pion mass. Combining this result with the value obtained from
a similar analysis using ${\cal O}^a_{41}$, we finally find
\begin{equation}
  \label{eq:xi1Kres}
  \langle\xi\rangle_K(4\,\mbox{GeV}^2) = 0.0275(5)(17)\,,\quad
  a_1^K(4\,\mbox{GeV}^2) = 0.0453(9)(29)\,.
\end{equation}
Again, the first error is statistical and the second systematic. The
magnitude of this quantity has been debated in the literature quite
recently, see the discussion in~\cite{Ball:2006wn}. The value we find
is more accurate than the sum rule estimate $a_1^K=0.05(25)$, but
fully compatible both in sign and in magnitude. The independent
investigation in~\cite{Boyle:2006pw} finds a value of
$a_1^K(4\,\mbox{GeV}^2) = 0.053(5)$ which is also compatible with our
finding.

% --------------------------------------------------------------------------
%
%  Summary
%
% --------------------------------------------------------------------------

\section{Summary\label{sec:summary}}
We have performed a calculation of the first two moments of the
distribution amplitudes of pseudoscalar mesons. In the case of the
pion we have found agreement with previous sum rule calculations and
phenomenological evaluations from experimental data. In the case of
the kaon we have also found agreement with previous sum rule
estimates, but with smaller statistical errors. The uncertainty due to
the continuum extrapolation has been estimated and found to be small.

\acknowledgments The numerical calculations have been done on the
Hitachi SR8000 at LRZ (Munich), on the Cray T3E at EPCC (Edinburgh)
under PPARC grant PPA/G/S/1998/00777 \cite{Allton:2001sk} and on the
APEmille and APEnext at NIC/DESY (Zeuthen). This work was
supported in part by the DFG under contract FOR 465 (Forschergruppe
Gitter-Hadronen-Ph{\"a}nomenologie) and in part by the EU Integrated
Infrastructure Initiative Hadron Physics (I3HP) under contract number
RII3-CT-2004-506078. W.S.~thanks the Physics Department of the
National Taiwan University for their hospitality and Jiunn-Wei Chen
for valuable remarks and discussions.

% --------------------------------------------------------------------------
%
%  Bibliography
%
% --------------------------------------------------------------------------

% --------------------------------------------------------------------------
% 
%  End of document
% 
% --------------------------------------------------------------------------

\end{document}